\documentclass[10pt,a4paper,twoside]{article}
\usepackage{amsmath}
\usepackage{bm} 
\usepackage{booktabs}
\usepackage{caption} 
\usepackage{dcolumn}
\usepackage{fancyhdr} 
\usepackage{graphicx}
\usepackage{hyperref}
\usepackage{latexsym}
\usepackage{xcolor}
\input{spp.dat}

\begin{document}

\title{\TitleFont Deformed Special Relativity using a Generalized\\ \lq t Hooft-Nobbenhuis Complex Transformation}


\author[*\negthickspace]{Jeffrey A. Q.~Abanto}
\affil[ ]{Astronomy Department, New Era University, Philippines}
\affil[*]{\corremail{jaqabanto@neu.edu.ph} }

\begin{abstract}
\noindent
A generalized form of \lq t Hooft-Nobbenhuis Complex space-time Transformation is applied on momentum space from which a new model of Deformed Special Relavity at Planck Scale is proposed. The model suggests an energy-dependent Planck's \lq\lq constant\rq\rq that vary in space and time and Quantum Mechanics as a low-energy approximation of perhaps a more fundamental theory at Planck Scale. 
\end{abstract}

\maketitle
\thispagestyle{titlestyle}

\section{Introduction}\label{sec:intro}
In a recent paper \cite{Aba1} by this author,  a generalized \lq{}t Hooft-Nobbenhuis space-time complex Transformation (GtHNT) was derived by modifying the Lorentz Transformation into its complex exponential form. It was shown that \lq{}t Hooft and Nobbenhuis' Transformation (tHNT) of space and time can be derived from a more general class of transformation that still preserve the Lorentz Invariance. This generalization gives more solid physical basis of tHNT since it can now be linked to the mathematical formalism of Relativity rather just treating the transformation as a mere mathematical tool to solve the Cosmological Constant Problem via a symmetry argument as initially intended by \lq{}t Hooft and Nobbenhuis. In this paper, our motivation is to explore the possiblity that GtHNT can be used in the mathematical formalism of Deformed Special Relativity to the extent that it can also be used to show the emergent nature of Quantum Mechanics. Recall that the suggested modification on Lorentz Transformation, $X'=\Lambda X$, is via the following complex transformation of the rapidity $\xi\rightarrow  i\alpha\xi$ and the Pure Lorentz Boost $
\Lambda$,
\begin{align}
&\Lambda\rightarrow L= \begin{pmatrix}
                       \cosh(i\alpha\xi)   &  -\sinh(i\alpha\xi)  \\                
		 -\sinh(i\alpha\xi) &  \cosh(i\alpha\xi)
                        \end{pmatrix} = \begin{pmatrix}
                       \cos(\alpha\xi)   &  -i\sin(\alpha\xi)  \\                
		 -i\sin(\alpha\xi) &  \cos(\alpha\xi)
                        \end{pmatrix} \\
&L \rightarrow \mathbf{A} L\mathbf{A}^{-1}=
                     \begin{pmatrix}
                     \cos(\alpha\xi)   &  -i\frac{1}{\alpha}\sin(\alpha\xi)  \\                
		-i\alpha\sin(\alpha\xi) &  \cos(\alpha\xi)
                        \end{pmatrix}
			=\mathbf{\Omega}^{*}
\end{align}
\noindent where $\alpha\ne \pm 1$ is a non-zero constant and $\mathbf{A}=\begin{pmatrix}
                       1 \; & \;0  \\                
		 0 \; & \;\alpha
                        \end{pmatrix}$ is a transformation matrix in terms of $\alpha$. Then, using a matrix generalization of Euler's Identity as suggested in \cite{Arg2007}, the complex Lorentz Boost $\mathbf{\Omega}^{*}$ can be written as follows;
\begin{equation}\label{genEu}
\mathbf{\Omega}^{*}=e^{-\xi \mathbf{\Phi}}
                       = \cos(\alpha\xi)\mathbf{I}-\frac{1}{\alpha}\sin(\alpha\xi)\mathbf{\Phi}
\end{equation}
\noindent where $\mathbf{I}=
                      \begin{pmatrix}
                       1   &  0  \\                
		0 & 1
                        \end{pmatrix} $ is the identity matrix and  $
\mathbf{\Phi}=
                      \begin{pmatrix}
                       0   &  i  \\                
		i\alpha^{2} & 0
                        \end{pmatrix} 
$ serves as an imaginary unit matrix.  Lorentz Invariance is still preserve since $\det\mathbf{\Omega^{*}}=1$. In \cite{Aba1}, we also noted a special case where we let $\mathbf{\Phi}=[i]$ and $\mathbf{I}=[1]$ become two 1 × 1 matrices, so that  $\mathbf{I}$ is the usual real unit and $\mathbf{\Phi}$ as the usual imaginary unit. If $\alpha= \pm 1$, it yields us the usual Euler Identity: $
e^{-i\xi}=\cos{\xi}- i\sin{\xi}=\varphi^{*}$, and gives us a GtHNT in terms of a complex number $\varphi^{*}$,
\begin{equation} 
t\rightarrow\varphi^{*} t\;\;\;\;\;\;\;\;\;\;\;\;\; x \rightarrow\varphi^{*} x
\end{equation} 
\noindent This is from a complex Lorentz Transformation $X\rq{}=\varphi^{*} X$ which still preserves Lorentz Invariance since $\det\varphi^{*}=det \begin{pmatrix}
                       \cos{\xi}   & \sin{\xi} \\                
		-\sin{\xi} & \cos{\xi}
                        \end{pmatrix} =1$. Lorentz Invariance will be violated if $\alpha\ne \pm 1$, i.e., $\Omega^{*}\rightarrow \tilde{\varphi^{*}}= \cos(\alpha\xi)-\frac{i}{\alpha}\sin(\alpha\xi)$ such that $\det\tilde{\varphi^{*}}=det \begin{pmatrix}
                       \cos{(\alpha\xi)}   & \frac{1}{\alpha}\sin{(\alpha\xi)} \\                
		-\frac{1}{\alpha}\sin{(\alpha\xi)} & \cos{(\alpha\xi)}
                        \end{pmatrix} =\cos^{2}(\alpha\xi)+\frac{1}{\alpha^{2}}\sin^{2}(\alpha\xi)=|\tilde{\varphi}|^{2}\ne 1$. In this paper, we explore the possibility that this result can be applied to energy dispersion relation from which a new model of Deformed Special Relativity(DSR) can be formulated. This is in line with the extension of Arbab\cite{Arbab_2010} of tHNT which includes the complex transformation of mass.

\section{New DSR}
 In retrospect, the seminal work done by Amelino-Camelia \cite{Am38} on what is now called as \lq\lq{}Deformed Special Relativity\rq\rq{} or \lq\lq{}Doubly Special Relativity\rq\rq{}, is an attempt to modify Lorentz Invariance at the Planck Scale. It uses the idea that the Planck energy or Planck length are fundamentally invariant and can be used to modify the momentum space at Planck scale. In simplest case, most DSR theories as enumerated in \cite{salesi2017} consider the invariant quantity $m^2 = E^{2} - p^2=(p^{0})^{2} - p^2$ and transformed it in various ways. One is putting in additional terms, 
\begin{equation}
m^2 = (p^{0})^{2} - p^{2}  - (p^{0})^{2}f(\frac{p}{M})
\end{equation} 
\noindent where  $M$ indicates  a mass scale characterizing the Lorentz breakdown. Others use momentum-dependent \lq\lq form factors'' $g=g(p)$ and $f=f(p)$, i.e., $
m^{2}=g^{2}(p)(p^{0})^{2}-f^{2}(p)p^{2}$, while in the work of Maguiejo and Smolin \cite{MaSmlet02} and Salesi et.al.\cite{salesi2017}, they used \lq\lq deformation function'' $F$,
\begin{equation}
m^2=F^{2}[(p^{0})^{2}-p^{2}]=F^{2}g_{\mu\nu}p^{\mu}p^{\nu}
\end{equation}
\noindent where $F$ is a function in terms of the Planck Length $\L_{p}$. Maguiejo and Smolin also proposed a momentum-dependent metric, $
 ds^{2}=g^{-2}(p)dt^{2}-f^{-2}(p)dl^{2}$ in their \lq\lq Rainbow Gravity'' theory \cite{Magueijo_2004} which may allow us to extend DSR to General Relativity and quantum gravity theories. In essence, most DSR theories conformally modifies the metric tensor $g_{\mu\nu}$ either via a scale-dependent transformation of momentum coordinates or fundamentally starts with a conformal transformation of the metric tensor and then derive
 the corresponding scale-dependent transformation of the momentum coordinates. However, it must also be considered that momentum coordinates should be expressed in terms of Planck constant and de Broglie wavelength that must be shown to be a low-energy approximation of the Planck length. Recent studies suggest that the Planck constant is a time-dependent variable at Planck scale \cite{time}. This must be incorporated also from which one can derived a modified form of Heisenberg Uncertainty Principle (HUP) at Planck scale. In this section, a simple derivation of a modified HUP as well as for a varying Planck \lq\lq constant\rq\rq will be shown by applying the GtHNT on the energy dispersion relation. The GtHNT implies the following transformation in the invariant quantity $m^{2}$,
\begin{align}
m^{2}=g_{\mu\nu}p^{\mu}p^{\nu}\rightarrow|\varphi|^{2}g_{\mu\nu}p^{\mu}p^{\nu}=g_{00}\tilde{E}\tilde{E}^{*}-g_{ij}\tilde{p}\tilde{p}^{*}
\end{align}
\noindent where we have set $\varphi=e^{i2\pi\chi}$ as a complex function for some scalar function $\chi$ and $|\varphi|^{2}\ne 1$ while $\tilde{E}=\varphi E$, $\tilde{E}^{*}=\varphi^{*} E$, $\tilde{p}=\varphi p^{i}$ and $\tilde{p}^{*}=\varphi^{*} p^{j}$. Evaluating $\tilde{E}$ and $\tilde{p}$, we use $i2\pi\varphi=\frac{\partial \varphi}{\partial \chi}$, $E=-\frac{\partial S}{\partial t}$, $p^{1}=\frac{\partial S}{\partial x}=p_{x}$ and inserting the imaginary number $i=\sqrt{-1}$, we have
\begin{equation}\label{optder2}
\tilde{E}=\varphi E=-\frac{i}{2\pi}(i2\pi\varphi) E =-\frac{i}{2\pi}\left(\frac{\partial\nonumber \varphi}{\partial \chi}\right)\left(-\frac{\partial S}{\partial t}\right)=\frac{i}{2\pi}\left(\frac{\partial \varphi}{\partial t}\right)\left(\frac{1}{f}\frac{\partial S}{\partial t}\right)
\end{equation}
\begin{equation}\label{optder3}
\tilde{p}_{x}=\varphi p_{x}=-\frac{i}{2\pi}(i2\pi\varphi)p_{x} =-\frac{i}{2\pi}\nonumber \left(\frac{\partial \varphi}{\partial \chi}\right)\left(\frac{\partial S}{\partial x}\right)=-\frac{i}{2\pi}\left(\frac{\partial \varphi}{\partial x}\right)\left(\lambda\frac{\partial S}{\partial x}\right)
\end{equation}
\noindent where $S$ is the classical action and the following variables were defined:
\begin{align}\label{frelen}
\frac{1}{f}=\frac{\partial t}{\partial \chi}\;\;\;\;\;\text{and} \;\;\;\;\lambda=\frac{\partial x}{\partial \chi}
\end{align}
\noindent Simplifying further, we define the variable,
\begin{align}\label{mpc}
\tilde{h}= \frac{1}{f}\frac{\partial S}{\partial t}=\lambda\frac{\partial S}{\partial x}=\frac{\partial S}{\partial \chi}
\end{align}
\noindent such that we yield the following equations: $\tilde{E}=\varphi E \;\;=i\tilde{\hbar}\partial_{t} \varphi$ and $\tilde{p}_{x}=\varphi{p}_{x}=-i\tilde{\hbar}\partial_{x}\varphi$, where $\partial_{x}$ and $\partial_{t}$ are the partial derivatives in space and time, respectively, and $\tilde{\hbar}=\frac{\tilde{h}}{2\pi}$. Since $\varphi$ is not an operator, it commutes with $E$ and $p_{x}$, thus we have the following eigenvalue equations: $
E\varphi=\hat{E}\varphi$  and $p_{x}\varphi= \hat{p}_{x}\varphi              
$, which gives us the following operator correspondence, 
\begin{align}
E\equiv i\tilde{\hbar}\partial_{t}=\hat{E}\;\;\;\;\;\;\;\;p_{x}\equiv -i\tilde{\hbar}\partial_{x}=\hat{p}_{x}
\end{align}   
\noindent that is very similar to Quantum Mechanics. Lastly, we use Eq. (\ref{mpc}) and set $S=S(\chi)$. Then integrating and setting the variable $\tilde{h}$ equal to a constant $\tilde{h}_{c}$, it will yield us
\begin{align}\label{chi1}
\chi=S/\tilde{h}_{c}+\textsf{k}
\end{align}
\noindent for some  integration constant $\textsf{k}$. This will transform $\varphi$ as follows:
\begin{align}
\varphi\rightarrow\varphi_{c}= Ae^{iS/\tilde{\hbar}}
\end{align}
\noindent which is similar to quantum probability amplitude where $A$ is a constant and $\tilde{\hbar}=\tilde{h}_{c}/2\pi$ is similar to reduced Planck constant.  All of the results above are equivalent to Quantum Mechanics if and only if the energy-dependent variable $\tilde{h}$ becomes constant and equal to the Planck constant. Furthermore, the idea that the quantum probability amplitude acts as the conformal factor of the metric tensor is in line with the earlier work of Dzhunushaliev \cite{Dzh1} where he suggested that the metric tensor can be considered to represent the microscopical state in a statistical system at Planck Scale. Also in the recent work of Isidro et.al.\cite{ric4} on Emergent Quantum Mechanics where they have shown that the State Vector or Wave Function is related to the conformal term of the metric tensor at Planck Scale. Thus, we consider here the suggestion of \lq t Hooft\cite{tHooft} that \lq\lq Local conformal symmetry could be as fundamental as Lorentz invariance,
and will guide us towards a complete understanding of Physics at the Planck scale''. We posit then the assumption that the conformal metric tensor,
\begin{equation}
\tilde{g}_{\mu\nu}=|\varphi|^{2}g_{\mu\nu}
\end{equation}  
\noindent applies at Planck Scale and from this, one can have a quantization of the usual metric tensor $g_{\mu\nu}$ where the conformal function $\varphi=\varphi(\tilde{h})$ becomes the quantum probability amplitude $\psi$ as the variable $\tilde{h}$ becomes the Planck's constant $h$, i.e., $\tilde{g}_{\mu\nu}\rightarrow \bar{g}_{\mu\nu}=|\psi|^{2}g_{\mu\nu}$. 
\section{Derivation of Quantum Kinematics}
 Relating now $f$ and $\lambda$ to the variable $\tilde{h}$ by using Eq.(\ref{mpc}), a modified de Broglie-Planck equations can be derived,
\begin{equation}\label{Pladeb}
E=\tilde{h}f\;\;\;\;\;\;\;\;\;\;\;p=\frac{\tilde{h}}{\lambda}
\end{equation}
\noindent Take note that the energy $E$ is the total energy of the system. What we wanted also is to define a minimum energy scale at Planck scale. For a field-free system with a constant total energy $E$ we have $S=\int Edt$, such that Eq. (\ref{chi1}) becomes,
\begin{align}
\chi=S/\tilde{h}_{c}+\textsf{k}=\frac{Et}{\tilde{h}_{c}}=\frac{E}{\tilde{h}_{c}f_{m}}=\frac{E}{E_{p}}
\end{align}  
\noindent where we set the integration contants to cancel each other, $E_{p}=\tilde{h}_cf_{m}$ and  $f_{m}=1/t$. If we define $E_{p}$ as the fundamental minimum energy, we can have a total energy $E$ in terms of $E_{p}$, i.e., $E=\chi E_{p}$. Now if at the Planck scale, the region in space (in one-dimension) that can be occupied by a fundamental particle is in N units of minimum length $L_{p}$, i.e., $
x=NL_{p}$, then by Eq.(\ref{frelen}), we have,
\begin{equation}
\lambda= \frac{\partial N}{\partial\chi}L_{p}+N\frac{\partial L_{p}}{\partial\chi }\;\;\;\;\;\;\;\;\;\frac{1}{f}=\frac{\partial N}{\partial\chi}\frac{L_{p}}{v}+\frac{\partial L_{p}}{\partial \chi}\frac{N}{v}
\end{equation}
\noindent 
\noindent where $v=\frac{\partial x}{\partial t}$. If $L_{p}$ is fundamentally invariant at Planck scale and $N$ is changing, we have,  
\begin{equation}
\lambda= \frac{\partial N}{\partial\chi}L_{p}\;\;\;\;\;\;\;\;\;\;\;\;\frac{1}{f}=\frac{\partial N}{\partial\chi}\frac{L_{p}}{v}
\end{equation}
\noindent Notice that the partial derivative in the equations is related to the energy density $\rho=\frac{E}{x} =\frac{\chi E_{p}}{N L_{p}}=\frac{\chi}{N}\rho_{p}$ where $\rho_{p}=E_{p}/L_{p}$ is the minimum energy density. For invariant $\rho_{p}$, $\frac{\partial N}{\partial \chi}=\frac{\rho_{p}}{\rho}-\frac{N}{\rho}\frac{\partial}{\partial \chi}\left(\frac{\chi}{N}\rho_{p}\right)$, which gives us,
\begin{align}
\lambda= \left[\frac{\rho_{p}}{\rho}-\frac{N}{\rho}\frac{\partial}{\partial \chi}\left(\frac{\chi}{N}\rho_{p}\right)\right ] L_{p}\;\;\;\;\;\;\;\;\;\;
f= \left[\frac{\rho_{p}}{\rho}-\frac{N}{\rho}\frac{\partial}{\partial \chi}\left(\frac{\chi}{N}\rho_{p}\right)\right ]^{-1} \frac{v}{L_{p}}
\end{align}
\noindent At this point, we note of the fact that the value of the energy density $\rho$ is inherently dependent on the measurement process as any measurement process will unavoidably add an energy into the system. At present, the way by which we do our measurement process is only at low energy resolution such that anything at Planck scale can only be observed in the order of Compton scale. If we have enough energy to increase the resolution of our measurement, such that we can send a single fundamental unit of energy $E_{p}$ to observe a single unit of fundamental length $L_{p}$, then $N=1=\chi$, $\rho=\rho_{p}$, which gives us, $\lambda = L_{p}$ and $f=\frac{v}{L_{p}}$ since $\frac{\partial \rho_{p}}{\partial\chi}=0$. Combining these results with Eq.(\ref{Pladeb}), we have,
\begin{equation}\label{mpladeb}
E= \tilde{h}\frac{v}{L_{p}}\;\;\;\;\;\;\;\;\;\;\;p=\frac{\tilde{h}}{L_{p}}
\end{equation}    
\noindent as the Planck scale equivalent of de Broglie-Planck equations which consider an invariant minimum length and a varying energy-dependent Planck \lq{}\lq{}constant". Furthermore, the limitation in our current measurement process will give rise to uncertainty. The total distance $x$ at Planck Scale has a corresponding uncertainty,
$\delta x= N \delta L_{p}$, where $\delta L_{p}$ is the uncertainty in measuring $L_{p}$. The uncertainty in measuring $L_{p}$ is brought about by the fact that the current nature of our measurement process will never have enough energy for it to have a resolution within the Planck scale. Substituting now $\delta L_{p}=N/\delta x$  to  Eq.(\ref{mpladeb}), it yields us
\begin{equation}
\delta E= \tilde{h}\frac{v}{\delta L_{p}}=\tilde{h}\frac{Nv}{\delta x},\;\;\;\;\;\;\;\;\;\;\;\delta p=\frac{\tilde{h}}{\delta L_{p}}=\frac{N\tilde{h}}{\delta x}
\end{equation}    
\noindent For $N>0$, we get a modified Heisenberg Uncertainty Principle, 
\begin{equation}
 \delta E\; \delta t >\tilde{h}\;\;\;\;\;\;\;\;\;\;\;\delta p \;\delta x >\tilde{h}
 \end{equation}
 \noindent where $\delta x$ and $\delta t=\delta x/v$ are the uncertainties in position and time, respectively, while $\delta E$ and $\delta p$ are the uncertainties in energy and momentum, respectively. At low-energy approximation where $\tilde{h}\rightarrow h$, it yields us the familiar Heisenberg Uncertainty Principle. In one sense, this is similar to Bohr\rq{}s Correspondence Principle but instead of the behavior of systems described by Quantum Mechanics reproducing Classical Physics, it is the behaviour of systems at Planck scale that is reproducing quantum-mechanical phenomena. Furthermore, it seems to prove that Quantum Mechanics is an incomplete theory that is emergent from a more fundamental theory at Planck Scale.  
\section{Conclusions}
We have shown that a new DSR model can be formulated by modifying the Lorentz Transformation using a generalize \lq t Hooft-Nobbenhuis spacetime complex transformation. In succeeding papers, we aim to show that quantum dynamics is related to the metric fluctuation that is happening at the Planck Scale.   
\section*{Acknowledgments}
\indent I would like to thank the NEU Board of Trustees and Administration and the University Research Center.
\bibliographystyle{spp-bst.bst}
\bibliography{bibfiledsr.bib}

\end{document}